# A Class of Errorless Codes for Over-loaded Synchronous Wireless and Optical CDMA Systems

P. Pad[1], F. Marvasti[1], K. Alishahi[2], S. Akbari[2]

*Abstract:* In this paper we introduce a new class of codes for over-loaded synchronous wireless and optical CDMA systems which increases the number of users for fixed number of chips without introducing any errors. Equivalently, the chip rate can be reduced for a given number of users, which implies bandwidth reduction for downlink wireless systems. An upper bound for the maximum number of users for a given number of chips is derived. Also, lower and upper bounds for the sum channel capacity of a binary over-loaded CDMA are derived that can predict the existence of such over-loaded codes. We also propose a simplified maximum likelihood method for decoding these types of over-loaded codes. Although a high percentage of the over-loading factor[3] degrades the system performance in noisy channels, simulation results show that this degradation is not significant. More importantly, for moderate values of $E_b/N_0$ (in the range of 6-10 dB) or higher, the proposed codes perform much better than the binary Welch bound equality sequences.

## I. Introduction

In a synchronous wireless[4] CDMA system with no additive noise, we can obtain errorless transmission by using orthogonal codes (Hadamard codes); we assume the number of users is less than or equal to the spreading factor (under or fully-loaded cases). In the over-loaded case (when the number of users is more than the spreading factor), such orthogonal codes do not exist; the choice of random codes creates interference that, in general, cannot be removed completely and creates errors in the Multi-User Detection (MUD) receiver [1-3].

---

[1] Advanced Communications Research Institute (ACRI) and Department of Electrical Engineering, Sharif University of Technology, Tehran, Iran.
[2] Department of Mathematical Sciences, Sharif University of Technology, Tehran, Iran.
[3] The percentage of the number of users divided by the number of chips minus 1, i.e., $(n/m - 1)$.
[4] In general, by wireless CDMA, we mean the signature codes (matrix) and the input data are binary $\{1, -1\}$; while for optical CDMA systems, the binary elements are $\{0,1\}$.

Likewise, for under-loaded optical CDMA systems, Optical Orthogonal Codes (OOC) [4-5] can be used. Unlike the connotation of the name of OOC, the optical codes are not really orthogonal, but by interference cancellation, we can remove the interference completely. However, for the fully and over-loaded cases, OOC's (with minimal cross-correlation value of $\lambda = 1$) do not exist and similar to the wireless CDMA, the choice of random codes creates interference that, in general, cannot be removed completely.

When the channel bandwidth is limited, the over-loaded CDMA may be needed. Most of the research in the over-loaded case is related to code design and Multi-Access Interference (MAI) cancellation to lower the probability of error. Examples of these types of research are pseudo random spreading (PN) [6-7], OCDMA/OCDMA (O/O) [8-9], Multiple-OCDMA (MO) [10], PN/OCDMA (PN/O) [11] signature sets, Serial and Parallel Interference Cancellation (SIC and PIC) [12-16]. The papers that discuss double orthogonal codes for increasing capacity [17-18] are actually non-binary complex codes (equivalent to $m$ phases for MC-OFDM) and are not really fair for comparison.

The codes with minimum Total Squared Correlation (TSC)[5] [20-22] maximize the channel capacity of a CDMA system when the input distribution is Gaussian [23]. But for binary input signals, the WBE codes do not necessarily maximize the channel capacity. Moreover, if the WBE codes are binary (BWBE), the optimality is no longer true. Another problem with WBE codes is that its ML implementation is impractical[6]. In our comparisons of our codes with WBE, we use iterative decoding methods with soft thresholding for WBE codes. For more details please refer to Section VI on simulation results.

None of the signatures and decoding schemes that have been proposed in the literature (including the BWBE) guarantee errorless communication in an ideal (high Signal-to-Noise Ratio (SNR) and without near-far effect) synchronous channel. In this paper, we plan to introduce Codes for Over-loaded Wireless (COW) and Codes for Over-loaded Optical (COO) CDMA systems [24] which guarantee errorless communication in an ideal channel and propose an MUD scheme for a special class of these codes that is Maximum Likelihood (ML). We will also compare these codes to BWBE and show that as the over-loading factor increases, the proposed COW/COO codes perform much better. As an example, for a signature length of 64, we have discovered such codes with an over-loading factor of about 62% that can be decoded practically in real time, which is also ML. However, we have proved the existence of codes with an over-loading factor of almost 156% that need to be discovered. The complexity of the decoding depends on the number of chips and the over-loading factor; but for a COW/COO code of size (64,104), the ML implementation is as simple as 8 look up tables of size 32. The implications of these findings are tremendous; it implies that using this system, we can accommodate 104 users for a spreading factor of 64 with low complexity ML decoding, which performs significantly better than BWBE in an AWGN channel (when $E_b/N_0$ is greater than 6 dB).

---

[5] Or equivalently, the Welch Bound Equality (WBE) [19] codes.
[6] There are some exceptions that are discussed in [29].

These codes are suitable for synchronous Code Division Multiplexing (CDM) in broadcasting, downlink wireless CDMA, and optical CDMA (assuming chip and frame synch). Alternatively, these codes can be used for the present downlink CDMA systems with much lower chip rate and hence significant bandwidth saving for the operating companies.

Using 64 chips, we have also derived an upper bound where the over-loading factor cannot be more than 320%. By trying to find bounds on the channel capacity in the absence of additive noise, we can, surprisingly, predict the existence of such codes.

Section II covers the necessary and sufficient conditions for errorless transmission in a noiseless over-loaded CDMA system along with methods for constructing large COW and COO codes with high percentage of over-loading factor. Two upper bounds for the number of users for a given signature length are presented in Section III. Channel Capacity evaluation for noiseless CDMA is discussed in Section IV. Methods for decoding are discussed in Section V. Simulation results and discussions are summarized in Section VI. Finally, conclusion and future work are covered in Section VII.

## II. Preliminaries-Channel Model

A synchronous CDMA system in an AWGN channel is modeled as

$$Y = \mathbf{C}AX + N,$$

where $\mathbf{C}$ is a matrix with signature columns with elements $\{1, -1\}$ or $\{0, 1\}$ depending on the application, $\mathbf{A}$ is a diagonal matrix with entries equal to the user received amplitude, $X$ is a binary user column vector with entries $\{1, -1\}$ or $\{0, 1\}$, $N$ is a white Gaussian noise with a covariance matrix of $\sigma^2 \mathbf{I}$ (where $\mathbf{I}$ is the identity matrix) and $Y$ is the received vector. In case of perfect power control, we can assume that $\mathbf{A} = \mathbf{I}$. Below we will discuss COW and COO codes.

### II.1 Codes for Over-loaded Wireless (COW) CDMA Systems

For developing COW and COO codes (matrices), we first discuss an intuitive geometric interpretation and then develop the codes mathematically. At a given time the multi-user binary data can be represented by an $n$-dimensional vector; these vectors can be interpreted as the vertices of a hyper-cube. Each user data is multiplied by a signature of $m$ chips long and finally their summation is transmitted. Thus, the transmitted $m$-tuple vectors are the multiplication of an $m \times n$ matrix (the columns are the signatures of different users) by the input $n$-dimensional vectors. Hence, the hyper-cube vertices are mapped onto points in an $m$-dimensional space ($m < n$). As long as the points in the $m$-dimensional space are distinct, the mapping is one-to-one and therefore, we can uniquely decode each received $m$-tuple vector at the receiver; on the other hand, if these $m$-tuple vectors are not distinct, the mapping is not one-to-one and the system is not invertible. Consequently, we look for codes that map the vertices of the $n$-dimensional hyper-cube to distinct points in the $m$-dimensional space. Most of the over-loaded codes discussed in the literature do not have this property and thus any MUD cannot be perfect. We coin the invertible codes, as mentioned in the introduction, as COW and COO codes for wireless and optical applications, respectively. We first develop systematic ways to generate COW codes and then extend it to COO codes.

**Lemma 1** We denote the vertices $\{1,-1\}^n$ of an $n$-dimensional hyper-cube with the set $\mathcal{V}$. The necessary and sufficient condition for the multiplication of a COW matrix $\mathbf{C}$ with elements of $\mathcal{V}$ to be a one-to-one transformation is $\text{Ker } \mathbf{C} \cap \{-1,0,1\}^n = \{0\}^n$, where $\text{Ker } \mathbf{C}$ is the null space of $\mathbf{C}$.

*Proof*: Let $X \in \text{Ker } \mathbf{C} \cap \{-1,0,1\}^n$. Then, $\mathbf{C}(2X) = 0$ and $2X$ is a $\{-2,0,2\}$-vector. Clearly, $2X = X_1 - X_2$, where $X_1$ and $X_2$ are $\{1,-1\}$-vectors. This implies that $\mathbf{C}X_1 = \mathbf{C}X_2$ and thus $X_1 = X_2$. Hence, $X = 0$ and the proof is complete. ∎

**Corollary 1** If $\mathbf{C}$ is a COW matrix, then
   a. A new COW matrix can be generated by multiplying each row or column of the matrix $\mathbf{C}$ by $-1$.
   b. New COW matrices can be generated by permuting columns and rows of the matrix $\mathbf{C}$.
   c. By adding an arbitrary row to $\mathbf{C}$, we obtain another COW matrix.

The proof is clear.

From Corollary 1, we can assume that all entries of the first row and the first column of a COW matrix are 1.

**Theorem 1** Assume that $\mathbf{C}$ is an $m \times n$ COW matrix and $\mathbf{P}$ is an invertible $k \times k$ $\{1,-1\}$-matrix, then $\mathbf{P} \otimes \mathbf{C}$ is a $km \times kn$ COW matrix, where $\otimes$ denotes the Kronecker product.

*Proof*: Clearly, $\mathbf{P} \otimes \mathbf{C}$ is a $\{1,-1\}$-matrix. Assume that $X$ is a $\{-1,0,1\}$-vector such that $(\mathbf{P} \otimes \mathbf{C})X = 0$. Then we have $(\mathbf{P}^{-1} \otimes \mathbf{I}_m)(\mathbf{P} \otimes \mathbf{C})X = 0$ and thus $(\mathbf{I}_k \otimes \mathbf{C})X = 0$. If $X^\mathrm{T} = [X_1^\mathrm{T} \ \cdots \ X_k^\mathrm{T}]^\mathrm{T}$, then we have $\mathbf{C}X_1 = \mathbf{C}X_2 = \cdots = \mathbf{C}X_k = 0$, where $X_i$'s are $n \times 1$ $\{-1,0,1\}$-vectors. Thus, by Lemma 1 we have $X_1 = X_2 = \cdots = X_k = 0$. Hence, no non-zero $\{-1,0,1\}$-vector is in the kernel of $\mathbf{P} \otimes \mathbf{C}$. Thus, $\mathbf{P} \otimes \mathbf{C}$ is a $km \times kn$ COW matrix. ∎

The existence of COW matrices with much higher percentage of the over-loading factors are given in the following theorem:

**Theorem 2** Assume $\mathbf{C}$ is an $m \times n$ COW matrix and $\mathbf{H}_2 = \begin{bmatrix} +1 & +1 \\ +1 & -1 \end{bmatrix}$. We can add $[(m-1)\log_3 2]$ columns to $\mathbf{H}_2 \otimes \mathbf{C}$ to obtain another COW matrix.

For the proof, refer to Appendix C.

**Note 1** $n/m \to \infty$ as $m \to \infty$.

This observation is a direct result of Theorem 2 since $n/m$ is of order $O(\log m)$. It implies that as the chip rate increases, the number of users grows much faster.

**Example 1** Applying Step 1 of the proof of Theorem 2 on a $2 \times 2$ Hadamard matrix, we first get a $4 \times 5$ COW matrix ($\mathbf{C}_{4\times 5}$) as shown in Table 1[7] (the $+$ sign represents 1 and the $-$ sign represents $-1$). By one more repetition, we find an $8 \times 13$ COW matrix ($\mathbf{C}_{8\times 13}$) depicted in Table 2. According to Theorem 1, $\mathbf{C}_{8\times 13}$ leads to a $64 \times 104$ COW matrix by the Kronecker product $\mathbf{H}_8 \otimes \mathbf{C}_{8\times 13}$ (where $\mathbf{H}_8$ is an $8 \times 8$ Hadamard matrix); this implies that we can have errorless decoding for 104 users with only 64 chips; i.e., more than %62 over-loading factor (we will introduce a suitable decoder for this code in Section VI). However, repetition of Theorem 2 for $\mathbf{C}_{8\times 13}$ shows the existence of a $64 \times 164$ COW matrix which implies an over-loading factor of about %156.

A fast algorithm for checking that a matrix is COW or not is given in Appendix B.

$$\begin{bmatrix} + & + & + & + & | & + \\ + & - & + & - & | & + \\ + & + & - & - & | & + \\ + & - & - & + & | & - \end{bmatrix}$$

Table 1. An example of $4 \times 5$ COW matrix-$\mathbf{C}_{4\times 5}$.

---

[7] Exhaustive search has shown that there are no $4 \times 6$ COW matrices.

$$\begin{bmatrix}
+ & + & + & + & + & + & + & + & + & + & + & + & + \\
+ & - & + & - & + & + & - & + & - & + & + & - & + \\
+ & + & - & - & + & + & + & - & - & + & - & + & - \\
+ & - & - & + & - & + & - & - & + & - & - & + & + \\
+ & + & + & + & + & - & - & - & - & - & + & - & + \\
+ & - & + & - & + & - & + & - & + & - & - & - & - \\
+ & + & - & - & + & - & - & + & + & - & - & - & + \\
+ & - & - & + & - & - & + & + & - & + & + & - & -
\end{bmatrix}$$

Table 2. An example of $8 \times 13$ COW matrix-$\mathbf{C}_{8\times 13}$.

**II.2. COO for Optical CDMA**

We would like to extend the results to optical CDMA, i.e., COO matrices.

**Theorem 3** If there is an $m \times n$ COW matrix for the wireless CDMA, then there is an $m \times n$ COO matrix for the optical CDMA.

*Proof*: Suppose **C** is an $m \times n$ COW matrix. By Corollary 1, we can assume that the entries of the first row of **C** are all 1. Now, we would like to prove that $\mathbf{D} = (\mathbf{J} + \mathbf{C})/2$ is a COO matrix, where **J** is the all 1 matrix. It is clear that **D** is a $\{0,1\}$-matrix. Assume $X \in \{-1,0,1\}^n$ and $\mathbf{D}X = 0$. This yields that $(\mathbf{J} + \mathbf{C})X = 0$ and thus $\mathbf{C}X = -\mathbf{J}X$. Because the entries of the first row of **C** are all 1, the first entry of $\mathbf{C}X$ is equal to the first entry of $\mathbf{J}X$. The above argument shows that the first entry of $\mathbf{J}X$ is 0. Thus $\mathbf{J}X = 0$. On the other hand $\mathbf{C}X = 0$ implies that $X = 0$, because **C** is a COW matrix. This shows that **D** is a COO matrix. ∎

**Corollary 2** Similar proof shows that, if we have a COO matrix which has a row with all 1's, then we will obtain a COW matrix by substituting the zeros of the COO matrix with $-1$.

**Example 2** As a special case, by Example 1 and Theorem 3, we also have a $64 \times 164$ COO matrix.

The theorems for COO matrices are similar to the previous theorems related to COW matrices. In addition, there are a few extra algorithms for the construction of COO matrices as described below.

**Theorem 4** If $\mathbf{D}$ is an $m \times n$ COO matrix, then $\mathbf{P} \otimes \mathbf{D}$ is also a $km \times kn$ COO matrix, where $\mathbf{P}$ is an invertible $k \times k$ {0,1}-matrix.

The proof is similar to the proof of Theorem 1.

**Corollary 3** If we set $\mathbf{P} = \mathbf{I}$ in the above theorem, then the generated COO matrices are sparse and have low weights that are suitable for optical transmission due to low power [4].

**Theorem 5** Suppose $\mathbf{A} = \mathbf{J} - \mathbf{I}$ is an $m \times m$ matrix, and $V_i = \left[ \underbrace{1 \cdots 1}_{2^i} \underbrace{0 \cdots 0}_{m-2^i} \right]^\mathrm{T}$, for $i = 0, \cdots, d$, where $d = \lfloor \log_2 m \rfloor - 2$. If $\mathbf{B} = [V_0 \ V_1 \ \cdots \ V_d]$, then $\mathbf{C} = [\mathbf{A}|\mathbf{B}]$ is an $m \times (m + d + 1)$ COO matrix.

*Proof*: Suppose $\mathbf{C}Z = 0$, where $Z$ is a $\{-1, 0, 1\}$-vector. Call the first $m$ entries of $Z$ by $X$ and the other $d + 1$ entries by $Y$. Hence, we have $\mathbf{A}X + \mathbf{B}Y = 0$ and this implies that $X = -\mathbf{A}^{-1}\mathbf{B}Y = -\left(\frac{\mathbf{J}}{m-1} - \mathbf{I}\right)\mathbf{B}Y = -\frac{\mathbf{J}}{m-1}\mathbf{B}Y + \mathbf{B}Y$. Obviously, $\mathbf{B}Y$ is an integer vector, thus it is sufficient to prove that $\frac{\mathbf{J}}{m-1}\mathbf{B}Y$ cannot be a non-zero integer vector. To show this we write $\frac{\mathbf{J}}{m-1}\mathbf{B}Y = \frac{1}{m-1}\begin{bmatrix} 1 & 2 & 4 & \cdots & 2^d \\ \vdots & \vdots & \vdots & \ddots & \vdots \\ 1 & 2 & 4 & \cdots & 2^d \end{bmatrix}$. Since $Y$ is a $\{-1, 0, 1\}$-vector then each entry of the vector $\mathbf{J}\mathbf{B}Y$ does not exceed $1 + 2 + \cdots + 2^d < m - 1$, thus $\frac{\mathbf{J}}{m-1}\mathbf{B}Y$ cannot be a non-zero integer vector. Now, suppose that $\mathbf{J}\mathbf{B}Y = 0$. If $Y = 0$, then $\mathbf{A}X = 0$. Since $\mathbf{A}$ is an invertible matrix, we conclude that $Z = 0$. Thus, assume that $Y \neq 0$. There exists an index $i$ such that $a_i 2^i + a_{i+1} 2^{i+1} + \cdots + a_d 2^d = 0$, $a_j \in \{-1, 0, 1\}$ for every $j$ and $a_i \neq 0$. This implies that $a_i$ is divisible by 2, a contradiction. ∎

**Example 3** Using Theorem 5, we get a $64 \times 69$ COO matrix with the structure discussed in the theorem.

In the next section we will try to find bounds on the number of users for a given spreading factor.

**Note 2** According to Lemma 1 if a matrix is COW, then any subset of its columns is also COW. This statement implies that if some of the users go inactive (we can assume that they are sending 0 instead of $\pm 1$), at the decoder we only need to know the active users (it is a common assumption in MUD [1-3]). Typically, in practical networks if a user becomes inactive, there are users in the queue that will grab the code. However, if we need a class of errorless codes that can detect inactive users for decoding, we must find the $\{1, -1\}$-matrices that operate injectively on $\{-1, 0, 1\}$-vectors. This is a topic we have covered in [27]. For COO matrices we do not have such problems since bit 0 is part of the transmitted data.

**III. Upper Bounds for the Percentage of Over-loading Factor**

Theorem 6 provides an upper-bound for the over-loading factor for a COW matrix.

**Theorem 6** If $\mathbf{C} = [c_{ij}]$ is a COW matrix with $n$ columns (users) and $m$ rows (chips), then

$$n \leq -m \left( \sum_{i=0}^{n} \frac{\binom{n}{i}}{2^n} \log_2 \frac{\binom{n}{i}}{2^n} \right)$$

where $\binom{n}{i} = \frac{n!}{i!(n-i)!}$.

*Proof:* Let the input multiuser data be defined by the random vector $X = [x_1, \ldots, x_n]^T$, where $x_i$'s are identically independent distribution random variables taking $-1, 1$ with probability $1/2$. Since $x_i$'s are independent, $H(X) = n$, where $H(X)$ is the entropy of $X$. Now, let the transmitted CDMA random vector be defined by $Y = \mathbf{C}X = [y_1, \cdots, y_m]^T$. For a given $j$, $1 \leq j \leq m$, the $n$ terms $c_{jk}x_k$, $k = 1, \cdots, n$ are independent random variables taking values $-1, 1$ with probability $1/2$. Hence

$y_j = \sum_{k=1}^{n} c_{jk} x_k$ is a binomial random variable with $H(y_j) = -\sum_{i=0}^{n} \frac{\binom{n}{i}}{2^n} \log_2 \frac{\binom{n}{i}}{2^n}$. We have $H(Y) \leq \sum_{j=1}^{m} H(y_j) = m\left(-\sum_{i=0}^{n} \frac{\binom{n}{i}}{2^n} \log_2 \frac{\binom{n}{i}}{2^n}\right)$. Now, because **C** is a COW matrix, then $X$ is also a function of $Y$ and thus $H(X) = H(Y) = n$, which completes the proof. ∎

**Note 3** In Appendix A, we estimate the entropy of $Y$ in another manner and derive a better upper bound. Fig. 1 shows this upper bound for the number of users versus the number of chips (spreading factor). This upper bound implies that with 64 chips, we cannot have a CDMA system with more than 268 users with errorless transmission. Ultimately, when the joint probabilities of all the $m$ elements of $Y$ are taken the maximum number of users with errorless transmission will be obtained. Using the above arguments, we can obtain similar upper bounds for COO codes.

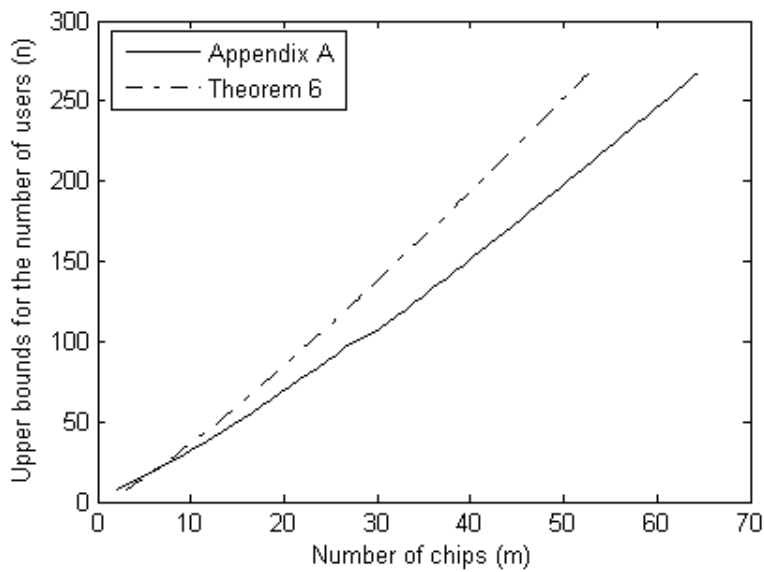

Fig. 1. The upper bounds for the number of users $n$ versus $m$ the number of chips (spreading factor). The dotted line is the bound from Theorem 6 while the solid line is the tighter bound derived from Appendix A.

**IV. Channel Capacity for Noiseless CDMA Systems**

In this section, we shall develop lower and upper bounds for the sum channel capacity [25] of a binary over-loaded CDMA with MUD when there is no additive noise [26]. The only interference is the over-loaded users. In this case, the channel is deterministic but not lossless. The interesting result is that the lower bound estimates a region for the number of users $n$ for a given chip rate $m$ such that COW or COO matrices exist. To develop the lower bound, we start by the following assumptions for the wireless case but results are also valid for the optical CDMA:

For a given $m$ and $n$, let $\mathcal{Q}_n = \{1, -1\}^n$ and $\mathcal{F}_{m,n}$ be the set of functions $f : \mathcal{Q}_n \to \mathbb{Z}_m$ defined by $f(X) = \mathbf{M}X$, where $\mathbf{M}$ is an $m \times n$ matrix with entries 1 and $-1$ and $X$ is the input multiuser vector as defined before with entries 1 and -1.

*Definition*: The sum channel capacity function C is defined as
$$C(m, n) = \max_{f \in \mathcal{F}_{m,n}} \log_2 |f(\mathcal{Q}_n)|$$
where $|\ |$ denotes the number of elements of the set. The above definition is equivalent to maximizing the mutual information $I(X, Y)$ which is equal to the output entropy (deterministic channel) over all the input probabilities and over all $m \times n$ $\mathbf{M}$ matrices ($X$ and $Y$ are binary $n \times 1$ and non-binary $m \times 1$ vectors, respectively).

**Lemma 2**
  (i) $C(m, n) \leq n$
  (ii) $C(m, n) \leq m \log_2(n + 1)$

*Proof*: (i) is trivial since $|f(\mathcal{Q}_n)| \leq |\mathcal{Q}_n| = 2^n$. For (ii) note that if $X \in \mathcal{Q}_n$ and $Y = [y_1 \cdots y_n]^T = f(X) = \mathbf{M}X$ then $y_j = \sum_{k=1}^n m_{jk} x_k$ is the sum of $-1$'s and $1$'s and can take $n + 1$ values. Hence, there are at most $(n + 1)^m$ possible vectors for $Y$.
∎

**Lemma 3** If $n$ is divisible by $m$, then $C(m, n) \geq m \log_2(n + m) - m \log_2 m$.

For the proof, refer to Appendix D.

To get tighter bounds than the ones given in Lemmas 2 and 3, we need the following theorems:

**Theorem 7 (Channel Capacity Lower Bound)**

$$C(m,n) \geq n - \log_2 A(m,n)$$

where

$$A(m,n) = \sum_{j=0}^{\lfloor \frac{n}{2} \rfloor} \binom{n}{2j} \left[\frac{\binom{2j}{j}}{2^{2j}}\right]^m.$$

For the proof, refer to Appendix E.

**Theorem 8 (Channel Capacity Upper Bound)**

$$C(m,n) \leq m\left(\frac{1}{2}\log_2 n + \log_2 \lambda\right) + 1$$

where $\lambda$ is the unique positive solution of the equation

$$(\lambda\sqrt{n})^m = m e^{\frac{-\lambda^2}{2}} 2^{n+1}.$$

For the proof, refer to Appendix F.

The plots of the channel capacity upper and lower bounds with respect to $n$ for a typical value of $m = 64$ is given in Fig. 2(a). Fig. 2(b) is a dual plot with respect to $m$ for a fixed value of $n = 220$. Plots of the channel capacity lower bounds with respect to $m$ and $n$ are given in Fig. 3. The plot of the lower bound from Lemma 3 is not shown since the bound is lower than the one from Theorem 7 (see Fig. 2(a)) for $n < 1000$, however, for large $n$ ($> 4000$), it is a better lower bound.

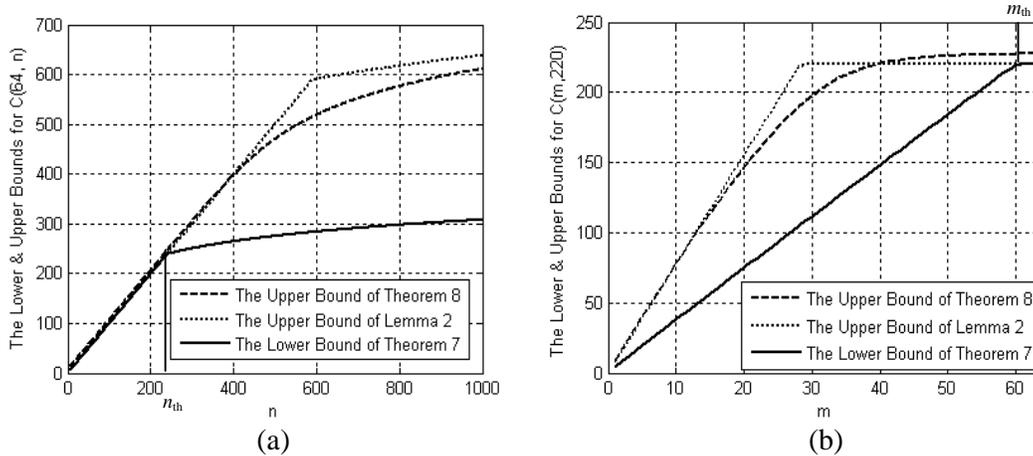

Fig. 2. Lower and upper bounds for the sum channel capacity with respect to: (a) the number of users $n$ for $m = 64$, (b) the chip rate $m$ for $n = 220$.

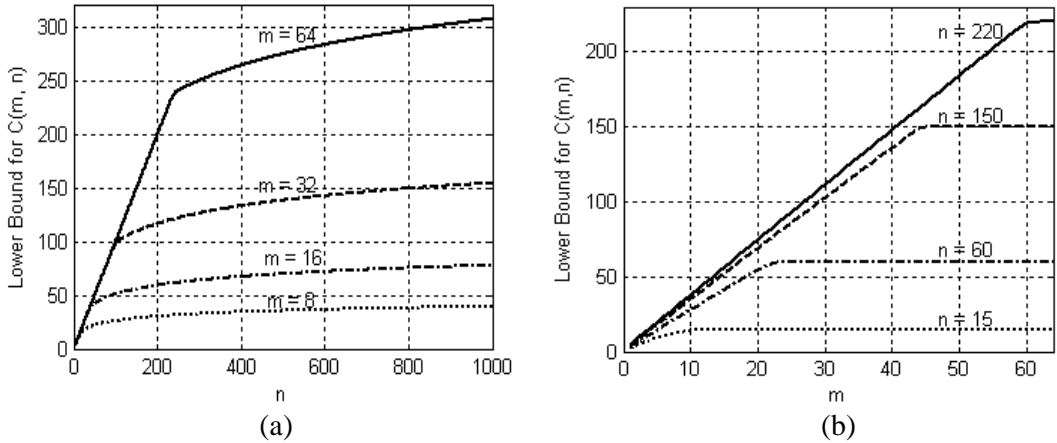

Fig. 3. Plots of channel capacity lower bounds for various $n$ and $m$: (a) lower bounds vs. number of users $n$ for a given chip rate $m$, (b) lower bounds vs. $m$ for a given $n$.

**Interpretation**: The lower bounds show interesting and surprising results. The lower bounds essentially show two modes of behavior. In the first mode, the lower bounds for the sum channel capacity (Fig. 2(a) and Fig. 3(a)) are almost linear with respect to $n$ for a given $m$, which implies the existence of codes that are almost lossless. Since we know that there exist COW (COO) codes that can achieve the sum channel capacity (number of users is equal to the sum channel capacity) without any error, the lower bound is very tight in this region. For small values of $m$ such as 4, we know that the maximum value of $n_{\max}$ such that a COW matrix exists is 5. The sum channel capacity lower bound for $m = 4$ is 4.21 bits, which is within a fraction of integer from 5. Also, for $8 \times 13$ COW matrix, the lower bound is 12.164 bits, which is again within a fraction of an integer from 13. We thus conjecture that the maximum number of users for a COW/COO matrix for $m = 64$ is around 239 from Fig. 2(a); right now our estimate from the simulations and upper bounds is an integer between 164 and 268.

After $n$ increases beyond a threshold value $n_{\text{th}}$ (Fig. 2(a)), the channel becomes suddenly lossy and enters the second mode of behavior. This loss is due to the fact that $2^n$ input points that are mapped to a subset of $(n + 1)^m$ points cannot find any empty space and a fraction of them get overlapped (no longer COW or COO condition).

Figs. 2(b) and 3(b) show another interesting behavior. Initially, the bound increases almost linearly with $m$ for a given $n$. This region is related to the case where the chip rate $m$ is much less than the number of users $n$. In our case, $n$ behaves like an amplitude or power, while $m$ behaves like frequency. As $m$ increases beyond a threshold ($m_{\text{th}}$ in Fig. 2(b)), the sum channel capacity remains almost constant since the capacity cannot be greater than $n$ (Lemma 2). In fact, $n$ is the supremum of the lower bound in this mode. This mode is the lossless case that predicts the existence of COW/COO codes.

The next section covers a practical ML algorithm for decoding a class of COW codes.

**V. Maximum Likelihood (ML) Decoding for a Class of COW/COO codes**

The direct ML decoding of COW codes is computationally very expensive for moderate values of $m$ and $n$. In this section, we prove two lemmas for decreasing the computational complexity of the ML decoders for a subclass of COW codes.

Suppose $\mathbf{D}$ is a COW/COO matrix and $Y = \mathbf{D}X + N$ is the received vector in a noisy channel. We wish to find a vector $\hat{X}$ $\{1, -1\}$ for wireless systems ($\{0,1\}$ for optical systems[8]) which is the best estimate of $X$ at the receiver. From now on we prove the lemmas for COW matrices but based on footnote[7], we can extend it to COO matrices.

**Lemma 4** Suppose $\mathbf{D}_{km \times kn} = \mathbf{P}_{k \times k} \otimes \mathbf{C}_{m \times n}$ where $\mathbf{P}$ is an invertible $\{1, -1\}$-matrix and $\mathbf{C}$ is a COW matrix. The decoding problem of a system with code matrix $\mathbf{D}$ can be reduced to $k$ decoding problems of a system with the code matrix $\mathbf{C}$.

*Proof*: Suppose $Y = \mathbf{D}X + N = (\mathbf{P} \otimes \mathbf{C})X + N$, where $N$ is the Gaussian noise vector with zero mean and auto-covariance matrix $\sigma^2 \mathbf{I}_{km}$ ($\mathbf{I}_{km}$ is the $km \times km$ identity matrix). Multiplying both sides by $\sqrt{k}(\mathbf{P}^{-1} \otimes \mathbf{I}_m)$, we have $Y' = \sqrt{k}(\mathbf{P}^{-1} \otimes \mathbf{I}_m)Y = \sqrt{k}(\mathbf{I}_k \otimes \mathbf{C})X + N'$ where $N' = \sqrt{k}(\mathbf{P}^{-1} \otimes \mathbf{I}_m)N$. This expression suggests that the first $m$ entries of $Y'$ depend on the first $n$ entries of $X$ and the first $m$ entries of $N'$; the second $m$ entries of $Y'$ depend on the second $n$ entries of $X$ and the second $m$ entries of the noise vector $N'$, and so on. Thus, retrieving the $(i-1)n+1, \ldots, i \cdot n$ entries of $X$ needs only the knowledge of the $(i-1)m+1, \ldots, i \cdot m$ entries of $Y'$, for $i = 1, \ldots, k$. Therefore, the decoding problem for $Y$ is decoupled to $k$ smaller decoding problems. ∎

---

[8] For a $\{0,1\}$-vector, we have $2Y - W = \mathbf{D}(2X - [1 \cdots 1]^{\text{T}}) + 2N$ where $W = \mathbf{D} \cdot [1 \cdots 1]^{\text{T}}$. Since $2X - [1 \cdots 1]^{\text{T}}$ is a $\{1, -1\}$-vector, ML decoding of $2Y - W$ is equivalent to ML decoding of $Y$.

In general, the ML decoding of the $Y'$ in Lemma 4 results in a sub-optimum decoder for $Y$. But if we suppose that the matrix $\mathbf{P}$ is a Hadamard one, since the matrix $\sqrt{k}(\mathbf{P}^{-1} \otimes \mathbf{I}_m)$ is a unitary matrix, the vector $N'$ is a Gaussian random vector with properties identical to $N$. Therefore, the ML decoding of $Y'$ is equivalent to ML decoding of $Y$. Since the ML decoding of $Y'$ is equivalent to the ML decoding of $k$ $m \times 1$ vectors, this implies a dramatic decrease in the computational complexity of the decoder in the over-loaded systems.

The following lemma introduces another method to significantly reduce the computational complexity of the decoder in over-loaded systems.

**Lemma 5** If a COW matrix $\mathbf{C}_{m \times n}$ is full rank, then the decoding problem for a system with code matrix $\mathbf{C}$ can be done through $2^{n-m}$ Euclidean distance measurements.

*Proof*: From part (b) of Corollary 1, we can always decompose the COW matrix $\mathbf{C} = [\mathbf{A}|\mathbf{B}]$ such that $\mathbf{A}$ is an $m \times m$ invertible square matrix. Assume $Y = \mathbf{C}X = \mathbf{A}X_1 + \mathbf{B}X_2$ where $X_1$ and $X_2$ are $m \times 1$ and $(n-m) \times 1$ vectors, respectively. Thus, $X_1 = \mathbf{A}^{-1}Y - \mathbf{A}^{-1}\mathbf{B}X_2$. Hence, we can search among $2^{n-m}$ possibilities of $X_2$ to find the vector $X_1$ that belongs to $\{1, -1\}^m$. In a noisy channel, we look for the specific $X_2$ that minimizes $\|(\mathbf{A}^{-1}Y - \mathbf{A}^{-1}\mathbf{B}X_2) - \text{sign}(\mathbf{A}^{-1}Y - \mathbf{A}^{-1}\mathbf{B}X_2)\|$, where $\| \; \|$ represents the Euclidean norm. The corresponding $X_1$ vector can be obtained by $X_1 = \text{sign}(\mathbf{A}^{-1}Y - \mathbf{A}^{-1}\mathbf{B}X_2)$, where $\text{sign}(Z)$ is obtained by substituting the positive entries of $Z$ by 1 and the negatives by $-1$. ∎

Similar to Lemma 4, Lemma 5 leads to significant decrease of the decoding complexity, but is not always optimum. Also, since the sign function maps a vector to the nearest $\{1, -1\}$-vector, it is not hard to show that if $\mathbf{A}$ is a Hadamard matrix, then the proposed method in Lemma 5 is an ML decoder.

Now, suppose that $\mathbf{C}_{m\times n} = [\mathbf{A}_{m\times m}|\mathbf{B}]$ and $\mathbf{D} = \mathbf{P}_{k\times k} \otimes \mathbf{C}$, where $\mathbf{A}$ and $\mathbf{P}$ are invertible matrices and $\mathbf{C}$ is a COW matrix. Combining Lemmas 4 and 5, we introduce a decoding scheme that has very low computational complexity, which is sub-optimum, in general. But if $\mathbf{A}$ and $\mathbf{P}$ are Hadamard matrices, the overall decoder is ML.

**Tensor Decoding Algorithm:** Suppose the received vector at the decoder is $Y = \mathbf{D}X + N$, where $N$ is the noise vector in an AWGN channel. The decoding algorithm is given below:

- *Step 1* Multiply both sides by $\mathbf{P}^{-1} \otimes \mathbf{I}_m$. We get $Y' = [{Y'_1}^T \cdots {Y'_k}^T]^T = (\mathbf{P}^{-1} \otimes \mathbf{I}_m)Y = (\mathbf{I}_k \otimes \mathbf{C})X + N' = (\mathbf{I}_k \otimes \mathbf{C})[X_1^T \cdots X_k^T]^T + N'$, where $Y'_i$ is the $(i-1)m+1, \ldots, i\cdot m$ entries of $Y'$ and $X_i$ is the $(i-1)n+1, \ldots, i\cdot n$ entries of $X$ for $i = 1, \ldots, k$.

- *Step 2* For each $i \in \{1, \ldots, k\}$, according to Lemma 5, multiply $Y'_i$ by $\mathbf{A}^{-1}$ and find the vector $\hat{X}_{2i}$ by minimizing $\|(\mathbf{A}^{-1}Y'_i - \mathbf{A}^{-1}\mathbf{B}\hat{X}_{2i}) - \text{sign}(\mathbf{A}^{-1}Y'_i - \mathbf{A}^{-1}\mathbf{B}\hat{X}_{2i})\|$ and set the vector $\hat{X}_{1i}$ to be equal to $\text{sign}(\mathbf{A}^{-1}Y'_i - \mathbf{A}^{-1}\mathbf{B}\hat{X}_{2i})$.

$\hat{X} = [\hat{X}_{11}^T \ \hat{X}_{21}^T \ \cdots \ \hat{X}_{1k}^T \ \hat{X}_{2k}^T]^T$ is the output of the decoder.

To see the power of this algorithm, let us take a CDMA system of size (64, 104) with the code matrix $\mathbf{D} = \mathbf{H}_8 \otimes \mathbf{C}_{8\times 13}$, where $\mathbf{H}_8$ denotes an $8 \times 8$ Hadamard matrix and $\mathbf{C}_{8\times 13}$ is the matrix shown in Table 2. Since $\mathbf{C}_{8\times 13}$ has an $8 \times 8$ Hadamard sub-matrix, the decoding of all the 104 users have a complexity of about $8 \times 32 = 2^8$ Euclidean distance calculation of 8-dimensional vectors. The decoder is also ML. This implies a drastic saving compared to the direct implementation of the ML decoder, which needs $2^{104}$ Euclidean distance calculation of 64 dimensional vectors.

In the next section, the COW codes with the proposed decoding method is simulated and compared to binary WBE and random codes.

## VI. Simulation Results

For studying the behavior of COW codes in the presence of noise, we consider three different CDMA systems in an AWGN channel. The first one is a system with the chip rate of $m = 64$ and $n = 72$ (64,72) users and the second one is of dimension (64,96) and the last one is (64,104). For each system, we compare three classes of codes: random, BWBE, and COW sequences. We use an iterative decoder with soft limiting[9] in the case of random and BWBE codes, which performs better than Parallel Interference Cancellation (PIC) with hard limiters [28]. For decoding COW codes, we apply the Tensor Decoding Algorithm (which is ML) discussed in the previous section. Note that we cannot use ML decoder for the BWBE[10] and random codes since their implementations are impractical. These decoding methods with the three different over-loading factors are compared with the orthogonal CDMA (Hadamard code of size $(64 \times 64)$), which performs the same as a synchronous binary PSK system- Figs. 4-6.

As seen in Fig. 4, for an over-loaded CDMA of size (64,72) for $E_b/N_0$ values less than 10 dB, the BWBE codes perform slightly better. But when $E_b/N_0$ increases beyond 10 dB, the Bit-Error-Rate (BER) of this system saturates. This phenomenon is due to the fact that the mapping of the BWBE code is not invertible. Thus when we use BWBE codes, we cannot decrease the BER lower than a threshold value even by increasing $E_b/N_0$ to infinity (or using any scheme of decoding). Since the mappings of COW codes are one-to-one and the proposed decoder is ML, the BER tends to zero as $E_b/N_0$ increases.

The simulation results of Fig. 4 are repeated in Figs. 5 and 6 for the other over-loaded COW codes (64,96) and (64,104), respectively. These figures highlight the fact that for higher over-loading factors, the COW codes with their simple ML decoding outperform other codes with iterative decoding. BWBE codes perform better than random codes due to its minimum TSC property, but the problem with such codes is that the interference cannot be cancelled totally and we cannot design optimum ML decoders due to their complexity. It is worth mentioning that in Fig. 6, although the system is about %62 over-loaded, the performance of COW codes is to within 3 dB of the orthogonal Hadamard fully-loaded CDMA, while the BWBE code has the same performance as the COW code for $E_b/N_0$ less 6 dB. But at higher $E_b/N_0$ values, the COW codes clearly outperform the BWBE Codes.

---

[9] F Marvasti, M Ferdowsizadeh, and P Pad , "Iterative synchronous and Asynchronous Multi-User Detection with Optimum Soft limiter" US Patent application number 12/122668 filed on 5/17/2008.
[10] There are some exceptions that are discussed in [29].

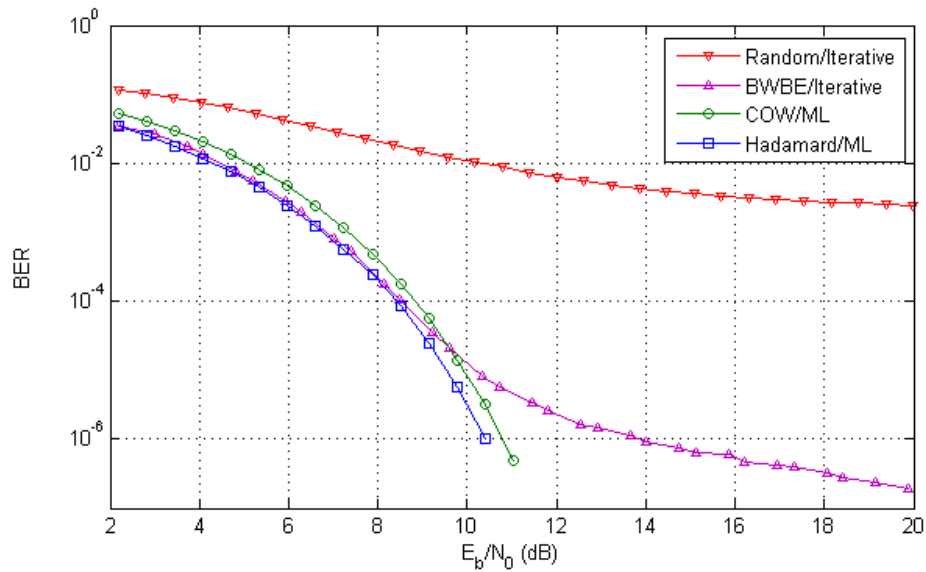

Fig. 4. Bit-error-rate versus $E_b/N_0$ for 3 classes of codes for a system with 64 chips and 72 users (for comparison, Hadamard codes of size $(64 \times 64)$ is also simulated).

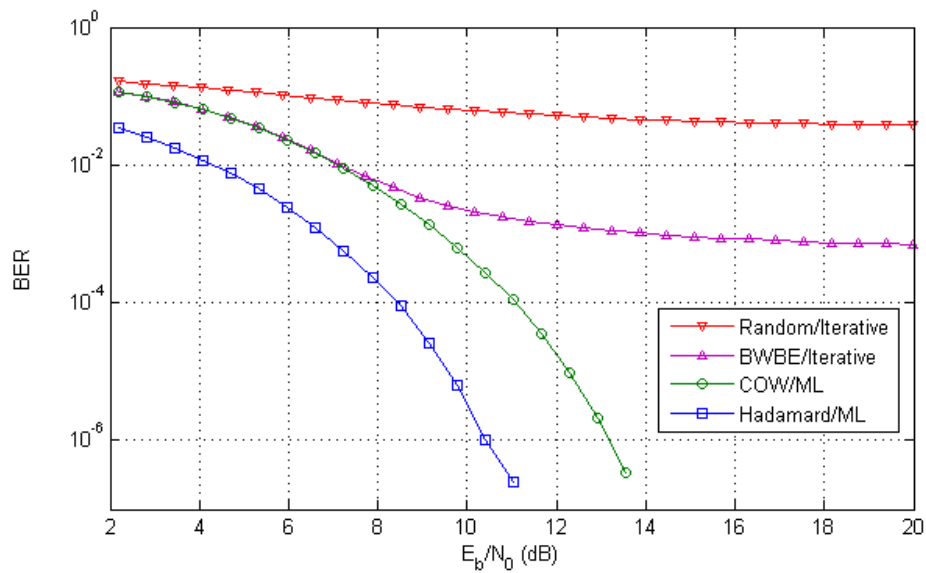

Fig. 5. Bit-error-rate versus $E_b/N_0$ for 3 classes of codes for a system with 64 chips and 96 users (for comparison, Hadamard codes of size $(64 \times 64)$ is also simulated).

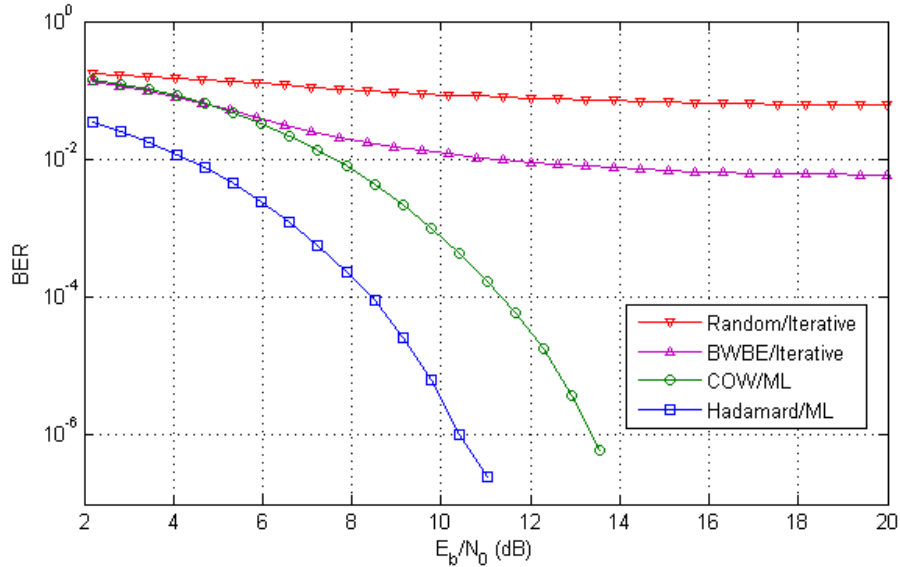

Fig. 6. Bit-error-rate versus $E_b/N_0$ for 3 classes of codes for a system with 64 chips and 104 users (for comparison, Hadamard codes of size $(64 \times 64)$ is also simulated).

## VII. Conclusion

In this paper, we have shown that there exists a large class of $(m \times n)$ codes $(m < n)$ that are suitable for over-loaded synchronous CDMA both for wireless and optical systems. For a given spreading factor $m$, an upper bound for the number of users $n$ has been found. For example for $m = 64$, the upper bound predicts a maximum of $n = 268$. A tight lower bound and an upper bound for the channel capacity of a noiseless binary channel matrix have been derived. The lower bound suggests the existence of COW/COO codes that can reach the capacity without any errors.

Mathematically, we have proved the existence of codes of size $(64,164)$. However, since the decoding of such over-loaded codes are not practical, we have developed codes of size $(64,104)$ that are generated by Kronecker product of a Hadamard matrix by a small matrix of size $(8,13)$. The decoding can be done by a look-up table of size 32 rows. These types of COW codes outperform BWBE codes and other random codes at high over-loaded factors and probability of errors of approximately less than $10^{-3}$.

We suggest for future work to get better upper bounds for the over-loaded CDMA systems, more practical codes at higher over-loading factors, and better decoding algorithms. Extensions to non-binary over-loaded CDMA, asynchronous CDMA, and channel capacity evaluations under fading and multipath environments are other issues that need further research. Also, to include fairness among users, we need to investigate the minimum distance of each COW/COO codes and its random allocation.

## Acknowledgment

We would like to sincerely thank the academic staff and the students of Advanced Communications Research Institute (ACRI), specially, Drs J.A. Salehi and M. Nasiri-Kenari, M. Ferdowsizadeh, A. Amini and V. Aref for their helpful comments.

## Appendix A

According to Theorem 3, if we find an upper bound for the number of users $n$ for a given number of chips $m$ for the COO codes, then this upper bound is also valid for COW codes. Suppose $C$ is a COO matrix, using part c of Corollary 1, we can add an all $1$ row as the $0^{th}$ row to $C$. If $X$ and $Y$ are two vectors in the proof of Theorem 6, then we have

$$H(Y) = H(y_0, y_1, y_2) + H(y_3, y_4 | y_0, y_1, y_2) + \cdots + H(y_{m-1}, y_m | y_0, y_1, \ldots, y_{m-2})$$
$$\leq H(y_0, y_1, y_2) + H(y_3, y_4 | y_0) + \cdots + H(y_{m-1}, y_m | y_0)$$
$$= H(y_0, y_1, y_2) + H(y_0, y_3, y_4) - H(y_0) + \cdots + H(y_0, y_{m-1}, y_m)$$
$$- H(y_0).$$

If we denote the maximum value of $H(y_0, y_1, y_2)$ over all possible configurations of the first and the second rows of $C$ by $H3$ and set $H1 = H(y_0)$, then we have $H(Y) \leq \frac{m}{2}(H3 - H1) + H1$. Since $C$ is a COO matrix, $H(Y) = n$. Consequently, $n \leq \frac{m}{2}(H3 - H1) + H1$. $H1$ is the entropy of a binomial r.v. and is depicted in the proof of Theorem 6.

For calculating $H3$, let $\begin{bmatrix} 1 & \cdots & & \cdots & 1 & 1 & \cdots & & \cdots & 1 \\ 1 & \cdots & & \cdots & 1 & 0 & \cdots & & \cdots & 0 \\ \underbrace{1 \cdots 1}_{a} & \underbrace{0 \cdots 0}_{b} & \underbrace{1 \cdots 1}_{c} & \underbrace{0 \cdots 0}_{n-a-b-c} \end{bmatrix}$ be $0^{th}$, $1^{st}$ and $2^{nd}$ rows of $C$. Thus, we have $H(y_0, y_1, y_2) = -\sum_{y_0, y_1, y_2} P(y_0, y_1, y_2) \log_2 P(y_0, y_1, y_2)$, where

$$P(y_0, y_1, y_2) = \sum_i \frac{\binom{n-a-b-c}{i}\binom{c}{y_0-y_1-i}\binom{b}{y_0-y_2-i}\binom{a}{y_2-y_0+y_1+i}}{2^n}.$$

## Appendix B

For testing a matrix to be a COW matrix, according to Lemma 1, the crudest algorithm is to check $3^n - 1$ vectors for the zero-vector. Now we introduce a better method to decrease this number down to $(3^{n-m} - 1)/2$. Assume that the matrix $C_{m \times n}$ is full rank (this is not a very restricting condition). Then there are $m$ columns of $C$ that form an invertible $m \times m$ matrix. Suppose these columns are the first $m$ columns of $C$ and coin the consructed invertible matrix by $A$ and the other columns by $B$. Thus, $C = [A|B]$. Using Lemma 1, we know that if $C$ is not a COW matrix, then there exists a $\{-1,0,1\}$-vector $X$ such that $\mathbf{C}X = 0$. Suppose $X^\mathrm{T} = [X_1^\mathrm{T}\ X_2^\mathrm{T}]$ such that $\mathbf{C}X = \mathbf{A}X_1 + \mathbf{B}X_2 = 0$. Thus $X_1 = -\mathbf{A}^{-1}\mathbf{B}X_2$. Hence to check that $\mathbf{C}$ is a COW matrix, we should search through different possibilities for $X_2$, i.e., $\{-1,0,1\}^{n-m}$ (except $\{0\}^{n-m}$) to see whether $-\mathbf{A}^{-1}\mathbf{B}X_2$ belongs to $\{-1,0,1\}^m$ or not. This needs $3^{n-m} - 1$ searches, but one half of these vectors are the negatives of the other half, thus we need only $(3^{n-m} - 1)/2$ searches.

**Appendix C**

We prove this theorem in 3 steps. Define $\mathbf{D} = \mathbf{H}_2 \otimes \mathbf{C}$ and $\mathcal{S} = \{\mathbf{D}X | X \in \{-1,0,1\}^{2n}\}$.

*Step 1*

An interesting observation is that if $Z \in \{1,-1\}^{2m}$ and $Z \notin \mathcal{S}$, then the matrix augmentation $[\mathbf{D}|Z]$ is a COW matrix. The proof of this step is trivial.

*Step 2*

We would like to prove that if $\mathcal{P} = Q + \{1,-1\}^{2m}$, where $Q$ is an arbitrary $2m \times 1$ integer vector, then $|\mathcal{S} \cap \mathcal{P}| \leq 2^{m+1}$. To show this, suppose that $Y \in \mathcal{S} \cap \mathcal{P}$. Then there exists a $\{-1,0,1\}$-vector $X_{2n \times 1} = [X_1^\mathrm{T}\ X_2^\mathrm{T}]^\mathrm{T}$, where $X_1, X_2 \in \{-1,0,1\}^n$ and $Y = \mathbf{D}X$.

$$Y = \mathbf{D}X = \begin{bmatrix} +\mathbf{C} & +\mathbf{C} \\ +\mathbf{C} & -\mathbf{C} \end{bmatrix} \begin{bmatrix} X_1 \\ X_2 \end{bmatrix} = \begin{bmatrix} \mathbf{C}X_1 + \mathbf{C}X_2 \\ \mathbf{C}X_1 - \mathbf{C}X_2 \end{bmatrix} \text{ and } Y_i = \mathbf{C}X_i \text{ thus } Y = \begin{bmatrix} Y_1 + Y_2 \\ Y_1 - Y_2 \end{bmatrix}.$$

Since there is a one-to-one correspondence between the set of vectors $[(Y_1 + Y_2)^\mathrm{T}\ (Y_1 - Y_2)^\mathrm{T}]^\mathrm{T}$ and the set of vectors $[Y_1^\mathrm{T}\ Y_2^\mathrm{T}]^\mathrm{T}$, the cardinality of the two sets are equal. Denote the $i^\text{th}$ entry of $Y_1$ by $(Y_1)_i$, thus we have $(Y_1)_i = \sum_{j=1}^n c_{ij}(X_1)_j \equiv$ (the number of nonzero entries of $X_1$) (mod 2). Hence the entries of $Y_1$ are either all odd or all even. Also this holds for $Y_2$. Since $Y \in \mathcal{P}$, then for every $i, 1 \leq i \leq m$, we have 
$$\begin{cases} (Y_1)_i + (Y_2)_i = Q_i \pm 1 \\ (Y_1)_i - (Y_2)_i = Q_{m+i} \pm 1 \end{cases}.$$

By an easy calculation the solutions of the above equations are
$$\begin{cases} (Y_1)_i = \frac{Q_i + Q_{m+i}}{2} \pm 1, (Y_2)_i = \frac{Q_i - Q_{m+i}}{2} \\ (Y_1)_i = \frac{Q_i + Q_{m+i}}{2}, (Y_2)_i = \frac{Q_i - Q_{m+i}}{2} \pm 1 \end{cases}.$$

The above solutions are in two categories. Category 1 consists of the solutions which have 2 choices for $(Y_1)_i$ and only one choice for $(Y_2)_i$, while category 2 consists of solutions with a single choice for $(Y_1)_i$ and 2 choices for $(Y_2)_i$.

Now, for the determination of $|\mathcal{S} \cap \mathcal{P}|$, first assume that all entries of $Y_1$ are even and $l$ entries of $Y_1$ have two choices. Hence, the number of $[Y_1^T \ Y_2^T]^T$ vectors are $2^l 2^{m-l} = 2^m$, because the $l$ corresponding elements in $Y_2$ have only one choice and the other $m - l$ elements in $Y_2$ have 2 choices.

The same assertion holds when all entries of $Y_1$ are odd. Thus, $|\mathcal{S} \cap \mathcal{P}|$ has at most $2^m + 2^m = 2^{m+1}$ elements.

*Step 3*

Now, suppose that we add $k$ columns to **D**, $k < \lceil (m-1) \log_3 2 \rceil$, and the resultant matrix, **E**, is a COW matrix. We wish to prove one can add another column to **E** to obtain a COW matrix with $2n + k + 1$ columns. Assume that $\mathbf{E} = [\mathbf{D}|\mathbf{F}]$, where $\mathbf{F} = [W_1|\cdots|W_k]$, and $W_i$ is a $2m \times 1$ vector, for $i = 1, \ldots, k$. Let $X \in \{-1, 0, 1\}^{2n+k}$, $X = [X_1^T \ X_2^T]^T$, where $X_1$ is a $2n \times 1$ vector and $X_2$ is a $k \times 1$ vector. Hence, $\mathbf{D}X_1 = \mathbf{E}X - \mathbf{F}X_2$. By Step 2 and the fact that $X_2$ has $3^k$ different possibilities, we have $|\mathcal{V} \cap \{1, -1\}^{2m}| = \sum_{X_2} |\{\mathbf{D} \cdot \{-1, 0, 1\}^{2n}\} \cap \{-\mathbf{F}X_2 + \{1, -1\}^{2m}\}| \leq 3^k 2^{m+1}$ where $\mathcal{V} = \{\mathbf{E}X | X \in \{1, -1\}^{2n+k}\}$.

Now, if $3^k 2^{m+1} < |\{1, -1\}^{2m}| = 2^{2m}$, then we can add another column to matrix **D** by applying Step 1. Thus, we can add at least $\lceil (m-1) \log_3 2 \rceil$ vectors to **D** and obtain a bigger COW matrix. ∎

**Appendix D**

Assume $\mathbf{H} = [H_1|\cdots|H_m]$ is an $m \times m$ Hadamard matrix. Let 
$$\mathbf{C} = \left[ \overbrace{H_1|\cdots|H_1}^{\frac{n}{m}} \Big| \cdots \Big| \overbrace{H_m|\cdots|H_m}^{\frac{n}{m}} \right]$$
be an $m \times n$ code matrix. If $X$ is a data vector, then

$CX = a_1 H_1 + \cdots + a_m H_m$, where for every $i$, $1 \le i \le m$, $a_i$ can take $\frac{n}{m} + 1$ different values. Thus, $CX$ can have $\left(\frac{n}{m} + 1\right)^m$ different values and thus its logarithm is a lower bound for the sum channel capacity. ∎

**Appendix E**

Pick $f \in \mathcal{F}_{m,n}$ randomly by choosing entries of the defining matrix of $f$ independently and uniformly from $\{1, -1\}$. For any vertex $X$ of the $n$-dimensional hyper-cube $\mathcal{Q}_n$, one has

$$E(|f^{-1}(f(X))|) = E\left(\sum_{X' \in \mathcal{Q}_n} 1_{f(X)=f(X')}\right) = \sum_{X' \in \mathcal{Q}_n} E\left(1_{f(X)=f(X')}\right)$$

$$= \sum_{X' \in \mathcal{Q}_n} P(f(X) = f(X'))$$

where $f^{-1}$, $1_{f(X)=f(X')}$, and P are the pre-image set, conditional if statement, and the probability function, respectively, and E is expectation over $f$.

If $X$ and $X'$ differ in $k$ places, then

$$P(f(X) = f(X')) = \begin{cases} 0 & \text{if } k = 2j+1 \\ \left(\dfrac{\binom{2j}{j}}{2^{2j}}\right)^m & \text{if } k = 2j \end{cases}$$

(Note that for $f(X)$ and $f(X')$ to be equal, all of their $m$ entries should be equal which are independent equiprobable events.) Combining the above equations, we get

$E(|f^{-1}(f(X))|) = \sum_{j=0}^{\lfloor n/2 \rfloor} \binom{n}{2j} \left(\dfrac{\binom{2j}{j}}{2^{2j}}\right)^m = A(m,n)$ and hence $E\left(\sum_{X \in \mathcal{Q}_n} |f^{-1}(f(X))|\right) = 2^n A(m,n)$. Thus, there exists an $f \in \mathcal{F}_{m,n}$ such that $\sum_{X \in \mathcal{Q}_n} |f^{-1}(f(X))| \le 2^n A(m,n)$. But if $|f(\mathcal{Q}_n)| = k$ and the pre-images of the $k$ values of $f(\mathcal{Q}_n)$ have cardinalities $n_1, \ldots, n_k$, then $\sum_{X \in \mathcal{Q}_n} |f^{-1}(f(X))| = \sum_{j=1}^{k} n_j^2$.

By Cauchy-Schwartz inequality: $2^n = \sum_{j=1}^{k} n_j \le \left(\sum_{j=1}^{k} n_j^2\right)^{1/2} \left(\sum_{j=1}^{k} 1\right)^{1/2} \le (2^n A(m,n))^{1/2} k^{1/2}$.

Thus, $k \geq 2^n/A(m,n)$ and $C(m,n) \geq \log_2 k \geq n - \log_2 A(m,n)$. ∎

## Appendix F

To prove the theorem, we need a classical inequality about large deviations of simple random walk:

Let $S_n = \delta_1 + \delta_2 + \cdots + \delta_n$, where $\delta_i$'s are independent and equal to $-1, 1$ with probability 1/2. For any $\lambda > 0$, from [30] we have $P(|S_n| > \lambda\sqrt{n}) \leq 2e^{\frac{-\lambda^2}{2}}$. Let $f(X) = \mathbf{M}X$ be the mapping with the maximum image size, i.e., $|f(\mathbf{Q}_n)| = 2^{C(m,n)}$. Pick $X \in \mathbf{Q}_n$ randomly with uniform distribution and let $Y = \mathbf{M}X = [y_1 \cdots y_m]^T$ for $j$, $1 \leq j \leq m$, $y_j = \sum_{k=1}^n m_{jk}x_k$ is a summation of $n$ independent random $-1, 1$'s (because of the randomness of $x_k$'s) and so according to the random walk property, $P(|y_j| > \lambda\sqrt{n}) \leq 2e^{\frac{-\lambda^2}{2}}$. This implies that if $\mathcal{R} = [-\lambda\sqrt{n}, \lambda\sqrt{n}]^m$, then $P(Y \notin \mathcal{R}) = P(\exists j\ 1 \leq j \leq m\ |y_j| > \lambda\sqrt{n}) \leq 2me^{\frac{-\lambda^2}{2}}$, which means that there are at most $2^{n+1}me^{\frac{-\lambda^2}{2}}$ points of $f(\mathbf{Q}_n)$ outside $\mathcal{R}$.

Now, notice that $|f(\mathbf{Q}_n) \cap \mathcal{R}|$ is at most equal to the number of integer points in $\mathcal{R}$ with all coordinates having the same odd or even parity as $n$ which is less than $(\lambda\sqrt{n})^m$. Combining these two facts, we get $2^{C(m,n)} = |f(\mathbf{Q}_n)| = |f(\mathbf{Q}_n) \cap \mathcal{R}| + |f(\mathbf{Q}_n) \cap \mathcal{R}^c| \leq (\lambda\sqrt{n})^m + 2^{n+1}me^{\frac{-\lambda^2}{2}} = 2(\lambda\sqrt{n})^m$.

The last equality comes from definition of $\lambda$ given in Theorem 8, which implies that $C(m,n) \leq m\left(\frac{1}{2}\log_2 n + \log_2 \lambda\right) + 1$. ∎

**RERENCES:**